\documentclass[twocolumn,prl,showpacs,preprintnumbers,amsmath,amssymb]{revtex4}

\usepackage{graphicx}
\usepackage{dcolumn}
\usepackage{bm}

\begin{document}

\textbf{Comment on `` Regional Versus Global Entanglement in Resonating-Valence-Bond States''}

In a recent Letter~\cite{reg}, Chandran and coworkers study the
entanglement properties of valence bond (VB) states. Their main result is
that VB states do not contain (or only an insignificant amount of) two-site
entanglement, 
whereas they possess multi-body entanglement. Two examples (``RVB gas and
liquid'') are given to illustrate this claim, which essentially comes from a
lower bound derived for spin correlators in VB states. While we do not
question that two-site entanglement is generically ``small'' for isotropic
VB states, we show in this Comment that (i) for the ``RVB liquid'' on the
square lattice, the calculations and conclusions of  
Ref.~\cite{reg} are incorrect. (ii) A simple analytical
calculation gives the exact value of the correlator for 
the ``RVB gas'', showing that the bound found in Ref.\onlinecite{reg} is tight. (iii)
The lower bound for spin correlators in 
VB states is equivalent
to a celebrated result of Anderson dating from more than 50 years ago.

The $SU(2)$ symmetry of VB states guarantees
that any two-spin reduced density matrix is a ``Werner
state'' fully characterized by a parameter $p$. The considered pair of spins
is entangled if $p>1/3$. Chandran {\it et
  al.} used quantum information concepts such as monogamy of entanglement
and quantum telecloning to obtain bounds on $p$. The number $p$ is simply
related to the 
correlator $\langle {\bf S}_i.{\bf  S}_j \rangle $ between these two spins
$1/2$ 
(${\bf S}={\bf \sigma}/2$, with ${\bf \sigma}$ Pauli matrices). We have
$\langle {\bf S}_i.{\bf  S}_j \rangle = -3/4 p$ (and not ``exactly equal to
the parameter $p$'' as stated in Ref.~\cite{reg}).

(i) The ``RVB liquid'' is  the equal amplitude superposition of all
nearest-neighbour (NN) VB coverings of a bipartite lattice. Exact
results can be obtained for small sizes $L$ of the square
$L\times L$ lattice. For $L=4$, we do not recover the value 
$p\simeq 0.2004$ of Ref.~\cite{reg}, but find
$p=0.4457579115872$ for periodic boundary conditions (BC) and
$p=0.2281115037$ in the interior of a sample with open BC. However, what really matters is the behavior for large $L$. Exact calculations are
difficult in this case, but Monte Carlo calculations are
possible~\cite{worm}. We computed the 
NN correlator $\langle {\bf S}_i.{\bf  S}_j
\rangle$  for large samples (up to $L=128$) on the square
lattice with periodic BC. The data of Fig. 1 shows that $p$ is larger than
$1/3$ in the thermodynamic limit (we find $p=0.3946(3)$, resulting in an
entanglement of formation of $\simeq 0.0215$). Therefore, the ``RVB liquid''
on the square lattice {\it does} 
possess two-site (NN) entanglement, contrary to the claim of Ref.~\cite{reg}.

(ii) The ``RVB gas'' is the equal amplitude
superposition of all {\it bipartite} VB coverings of a bipartite
lattice. This is in fact the projection into the singlet sector of
the (magnetically ordered) {\it N\'eel state} on this
lattice. This observation can be used to calculate $p$ exactly. The total
spins ${S}_A$ and $S_B$ on  
sublattices A and B are maximal, couple antiferromagnetically and form a
singlet (total spin $S=0$). For a 
system of $2N$ spins, ${ S}_A={ S}_B=N/2$. One then
easily obtains that $\langle 
  {\bf S}_i.{\bf  S}_j \rangle=-1/4-1/(2N)$ if $i$ and $j$ belong to different
  sublattices. The equivalent exact result $p=1/3+2/(3N)$ shows that the
  telecloning bound $p\leq 1/3+2/(3N)$ is tight. Two-site entanglement is 
  therefore present in any finite ``RVB gas'' and vanishes only in the
  thermodynamic limit.
\begin{figure}[h]
\includegraphics*[height=2.7cm]{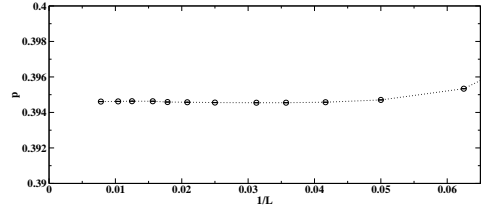}
\caption{Werner parameter $p$ as a function of inverse linear system
  size $1/L$ for the square lattice ``RVB liquid''.}
\end{figure}

(iii) The telecloning bound on $p$ in Ref.~\cite{reg} reproduces an inequality 
found by Anderson~\cite{Anderson}, who derived a
lower bound 
for the energy of antiferromagnetic spin models.
Take a spin at
site $i$, separated by any distance from a 
number $z$ of symmetry-equivalent spins $j$: $\langle {\bf S}_i.{\bf  S}_j 
\rangle$ (as well as $p$) is identical for all $z$ spins at sites $j$. In this
case, the telecloning bound is $p\leq 1/3+2/(3z)$ or equivalently $\langle
{\bf  S}_i.{\bf  S}_j \rangle \geq -1/4-1/(2z)$, the result derived by
Anderson.  His result (of variational nature) on correlators in a given
state is very general: it holds also for states other than singlets, is
independent of any Hamiltonian and can be refined further (see e.g. Ref.~\cite{Valenti}).  

In conclusion, the bound obtained with quantum
information techniques~\cite{reg} has been familiar in the condensed matter
context for a long time. Nevertheless, it is interesting to see that it can
be derived in a totally different framework. For the two examples chosen in
Ref.~\onlinecite{reg}, typical condensed matter methods allowed us to
provide in one case an  exact solution, and to show that the results of
Ref.~\onlinecite{reg} are incorrect in the other one. 

We thank M. Mambrini for useful discussions, and D. Kaszlikowski and
V. Vedral for correspondence. 

Fabien Alet, Daniel Braun \\
Laboratoire de Physique Th\'eorique, UMR CNRS 5152, Universit\'e Paul Sabatier, 31062 Toulouse, France 

Gr\'egoire Misguich \\
Institut de Physique Th\'eorique, URA CNRS 2306, CEA Saclay, 91191 Gif sur
Yvette, France


\vspace*{-1em}
\begin{thebibliography}{99}
\vspace*{-1em}
\bibitem{reg} A. Chandran {\it et al.}, Phys. Rev. Lett. {\bf 99}, 170502 (2007).
\bibitem{worm} We generalized the algorithm of A.W. Sandvik and R. Moessner,
  Phys. Rev. B {\bf 73}, 144504 (2006) to account for VB overlap properties.
\bibitem{Anderson} P.W. Anderson, Phys. Rev. {\bf 83}, 1260 (1951).
\bibitem{Valenti} R. Tarrach and R. Valent\'i, Phys. Rev. B {\bf 41}, 9611 (1990).
\end{thebibliography}
\end{document}